\documentclass[aps,prb,twocolumn,superscriptaddress,amsmath,amssymb]{revtex4}

\usepackage{graphicx}
\usepackage{dcolumn}
\usepackage{graphics}
\usepackage{amssymb}
\usepackage{bm}
\usepackage[ansinew]{inputenc}
\usepackage{placeins}

\begin{document}

\title{Orbitally defined field-induced electronic state in a Kondo lattice}

\author{G. G. Lesseux}
\affiliation{Instituto de F\'{\i}sica "Gleb Wataghin", Universidade Estadual de Campinas, 13083-859, Campinas, SP, Brazil}

\author{H. Sakai}
\affiliation{Advanced Science Research Center, Japan Atomic Energy Agency, Tokai, Ibaraki, 319-1195, Japan}

\author{T. Hattori}
\affiliation{Advanced Science Research Center, Japan Atomic Energy Agency, Tokai, Ibaraki, 319-1195, Japan}

\author{Y. Tokunaga}
\affiliation{Advanced Science Research Center, Japan Atomic Energy Agency, Tokai, Ibaraki, 319-1195, Japan}

\author{S. Kambe}
\affiliation{Advanced Science Research Center, Japan Atomic Energy Agency, Tokai, Ibaraki, 319-1195, Japan}

\author{P. L. Kuhns}
\affiliation{National High Magnetic Field Laboratory, Florida State University, Tallahassee-FL, 32310, U.S.A.}

\author{ A. P. Reyes}
\affiliation{National High Magnetic Field Laboratory, Florida State University, Tallahassee-FL, 32310, U.S.A.}

\author{J. D. Thompson}
\affiliation{Los Alamos National Laboratory, 87545, Los Alamos, NM, USA}

\author{P. G. Pagliuso}
\affiliation{Instituto de F\'{\i}sica "Gleb Wataghin", Universidade Estadual de Campinas, 13083-859, Campinas, SP, Brazil}

\author{R. R. Urbano}
\affiliation{Instituto de F\'{\i}sica "Gleb Wataghin", Universidade Estadual de Campinas, 13083-859, Campinas, SP, Brazil}

\date{\today}

\begin{abstract}
CeRhIn$_{5}$ is a Kondo-lattice prototype in which a magnetic field \textit{B}$\bf{^{\ast}\simeq}$ 30 T induces an abrupt Fermi-surface (FS) reconstruction\cite{Jiao2015_673,Moll2015_6663} and pronounced in-plane electrical transport anisotropy\cite{Ronning2017_313} all within its antiferromagnetic state. Though the antiferromagnetic order at zero field is well-understood, the origin of an emergent state at \textit{B}$^{\ast}$ remains unknown due to challenges inherent to probing states microscopically at high fields. Here, we report low-temperature Nuclear Magnetic Resonance (NMR) measurements revealing a discontinuous decrease in the $^{115}$In formal Knight shift, without changes in crystal or magnetic structures, of CeRhIn$_{5}$ at fields spanning \textit{B}$^{\ast}$. We discuss the emergent state above \textit{B}$^{\ast}$ in terms of a change in Ce's 4\textit{f} orbitals that arises from field-induced evolution of crystal-electric field (CEF) energy levels. This change in orbital character enhances hybridisation between the 4\textit{f} and the conduction electrons (\textit{c.e.}) that leads ultimately to an itinerant quantum-critical point at \textit{B}$\bf{_{c0} \simeq}$ 50 T.
\end{abstract}

\maketitle

Development of the peculiar electronic state above \textit{B}$\bf{^{\ast}}$  in CeRhIn$_{5}$ is signaled clearly in quantum oscillations,\cite{Jiao2015_673} magnetoresistance,\cite{Moll2015_6663,Jiao2015_673} magnetostriction\cite{Rosa2019_016402} but not in specific heat.\cite{Jiao2019_045127} The lack of a detectable specific heat anomaly suggests that \textit{B}$^{\ast}$ may not reflect a well-defined phase transition but a crossover from one state to another\cite{Rosa2019_016402} where not only the Fermi surface reconstructs from small-to-large\cite{Jiao2015_673} but also in-plane anisotropy develops in electrical resistivity.\cite{Ronning2017_313} Qualitatively, these responses could be consistent with a field-induced change in crystal and or magnetic structure from below to above \textit{B}$^{\ast}$ -- a distinctly plausible interpretation that could be tested straightforwardly by a diffraction measurement if \textit{B}$^{\ast}$ were sufficiently low to be accessible in neutron or x-ray experiments. Even if such measurements could be made at fields to 30 T and higher, experiments point to a more complex picture, with similarities to other correlated electron systems. Electrical resistivity studies reveal a hysteretic transition at \textit{B}$^{\ast}$ that was interpreted intially to reflect the formation of a density wave, analogous to that found in correlated copper-oxide materials.\cite{Moll2015_6663} More recent studies are even more surprising:\cite{Ronning2017_313} when an applied field is tipped about 20$^0$ from the tetragonal \textit{c}-axis toward in-plane perpendicular directions, there is a striking inequivalence of electrical resistivity for current flow along each pair of orthogonal crystallographic directions. This unexpected in-plane symmetry breaking is consistent with a proposed strong $XY$ nematic susceptibility that is similar to but distinct from Ising-nematicity that is found in high-$T_c$ copper oxide,\cite{Ando2012_137005,Kivelson1998_550} iron-pnictide\cite{Fernandes2013_137001,Chuang2010_181} and correlated ruthenate materials.\cite{Borzi2007_214} 

Evidence for all the changes in electronic properties at \textit{B}$^{\ast}$ and their weak coupling to the crystal lattice\cite{Ronning2017_313,Rosa2019_016402} appears only within the magnetically ordered state of CeRhIn$_{5}$. In this limit, partially Kondo-compensated Ce moments order below $T_{\rm N}$ = 3.8 K in a spin-spiral structure with ordering wave-vector $\bm{Q} = (0.5, 0.5, 0.297)$ and moments in the tetragonal plane.\cite{Fobes2018_NP} This structure, however, is unstable against modest applied pressure\cite{Aso2009_0737703} or in-plane applied magnetic field.\cite{Takeuchi2001_877,Raymond2007_242204,Fobes2018_NP} The near degeneracy of accessible orders in CeRhIn$_{5}$ supports the possibility that a field of 30 T could change the nature of magnetism at \textit{B}$^{\ast}$, but with little change in entropy or susceptibility. What might underlie the emergence of the new electronic state above \textit{B}$^{\ast}$ and a change in magnetic character, if this indeed happens, are fundamental questions raised by recent discoveries in CeRhIn$_{5}$ and are relevant more broadly to the physics of a Kondo lattice.

\begin{figure*}
\includegraphics[keepaspectratio,width=17.2cm]{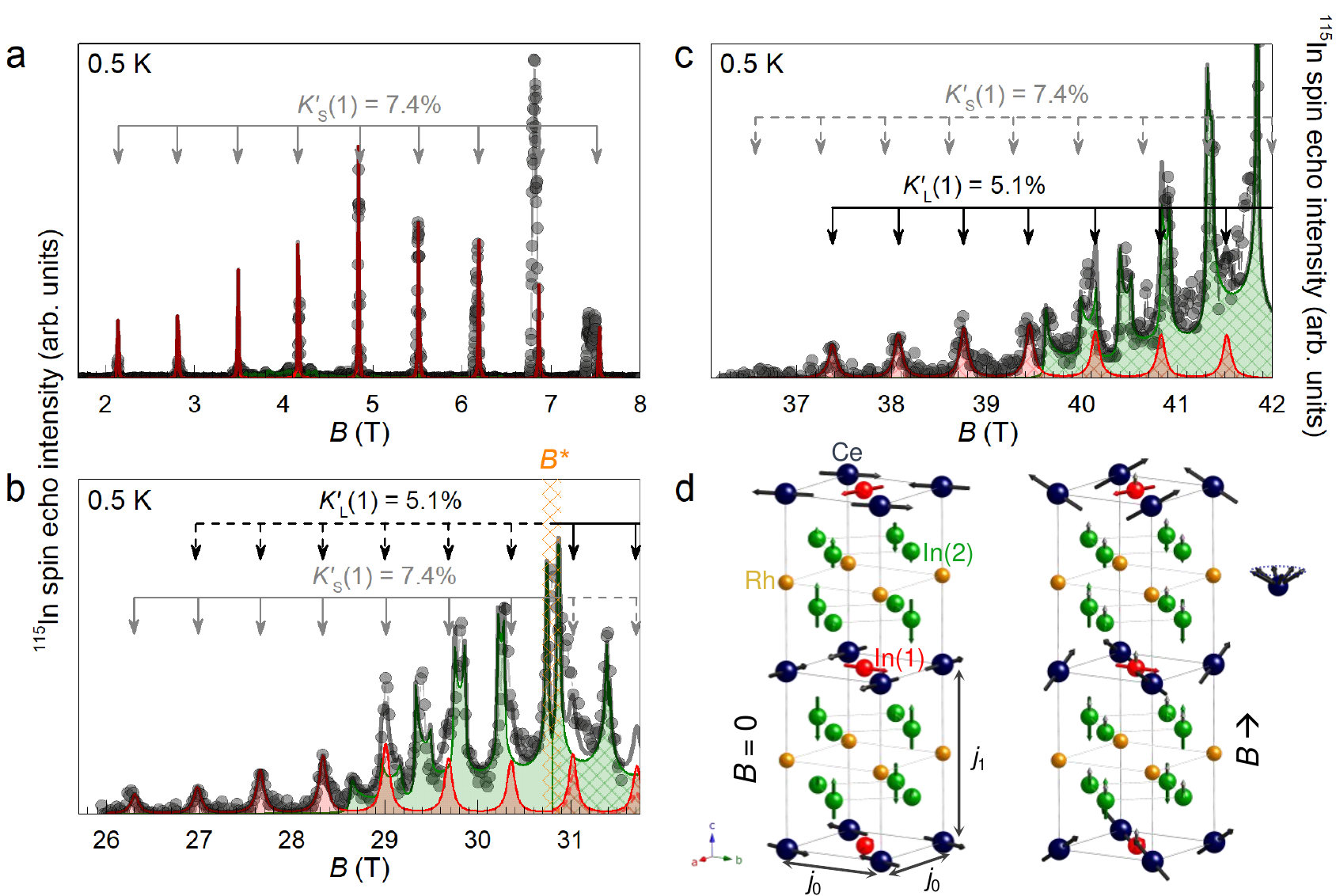}
\caption{High field $^{115}$In NMR spectra for distinct regions of the \textit{B} along-\textit{c} versus \textit{T} phase diagram of CeRhIn$_5$. \textbf{a}, \textbf{b} and \textbf{c} are In(1) and In(2) NMR spectra at 0.5 K for excitation frequencies 48.5 MHz, 290.65 MHz and 393.6 MHz respectively. The shaded magenta areas are simulations for the In(1) equidistant transitions, and the green area indicates a simulation for the incommensurate In(2) pattern (see Supplemental Material). The solid gray curve is the overall In-NMR simulated spectrum that includes both contributions. The non-hashed and hashed areas indicate the formal Knight shifts of In(1) and In(2) below ($K_{\rm{S}}'(1)$ = 7.4(1)\%, $K_{\rm{S}}'(2)$ = 1.5(2)\%) and above ($K_{\rm{L}}'(1)$ = 5.2(1)\%, $K_{\rm{L}}'(2)$ = 1.1(2)\%) the crossover at $B^{\ast} \sim$ 30.8 T. The crossover field is indicated by a vertical orange dashed strip in b). Light gray symbols are data. The vertical gray and black arrows indicate the expected In(1) transitions for formal Knight shifts $K_{\rm{S}}'(1)$ = 7.4\% and $K_{\rm{L}}'(1)$ = 5.2\%, respectively. The solid and dashed arrows indicate whether In(1) transitions were observed (solid) or not (dashed). The difference between gray and black arrows indicates a change in the shift $\Delta K'(1)$ for fields above $B^{\ast} \sim$ 30 T, but the line-shape and width of the transitions remain similar across $B^{\ast}$. Although smaller, there also is a change in $K'(2)$ as well, which follows the same trend as $K'(1)$ and is discussed in the text. To assure confidence in these measurements in the AF$_S$ and AF$_L$ phases, we measured a spectrum at 290.65 MHz while sweeping the field up and at 393.6 MHz in a down-field sweep. Results were reproducible. An Al-NMR signal (not shown) was used as a field marker. The probe used to acquire the 48.5 MHz spectrum shows extrinsic NMR signals from $^{207}$Pb and/or $^{209}$Bi present in the solder and coaxial cable in the NMR circuit visible at 0.5 K, $B \sim$  6.8 T and 7.5 T. \textbf{d} CeRhIn$_5$ crystallographic structure. The Ce, Rh, In(1) and In(2) sites are indicated in dark blue, yellow, red and green, respectively. In the left figure, we also show the magnetic structure (black arrows at the Ce site) for zero field and, in the right figure, how the magnetic structure evolves to a conical configuration by applying a magnetic field along the $c$-axis. The respective hyperfine (internal) fields for In(1) and In(2) sites are indicated by red and green arrows, respectively. The field induced hyperfine (internal) field at In(1) and In(2) sites are indicated by gray arrows in the right figure.}
\label{Fig1}
\end{figure*}

With its sensitivity to local spin, charge and lattice degrees of freedom,\cite{Bauer2011_6857, Seo2013_120} NMR is a powerful tool to probe the evolution of complex electronic states in correlated electron materials at very high magnetic fields.\cite{Wu2011_191,Sakai2014_236401,Berthier2017_331} Figure \ref{Fig1} presents the $^{115}$In NMR spectra ($I$ = 9/2) from two inequivalent sites of our CeRhIn$_5$ single crystal with $B$ applied along the $c$-axis at 0.5 K below $T_{\text{N}}(B)$. Each Ce atom is surrounded by four tetragonally coordinated In(1) and eight In(2) atoms with local orthorhombic symmetry (Fig. \ref{Fig1}d and Supplemental Material). At low fields (Fig. \ref{Fig1}a), there are 9 equally-separated transitions associated with In(1) NMR. In contrast, the lower relative intensities of the In(2) NMR signal are a consequence of spectral broadening due to a distribution of internal fields arising from an oscillating hyperfine (internal) field $B_{int}^{\| c}(2)$ associated with $c$-axis incommensuration of the spin-spiral magnetic structure\cite{Curro2000_R6100,Curro2006_173} (see Figure \ref{Fig1}d and Supplemental Material).

At low-fields and well below $T_{\rm N}$, a hyperfine field of $B_{int}^{\bot c}(1) = 0.17$ T lies in the Ce-In(1) ($ab$-) plane and rotates between the adjacent layers with the incommensurate pitch of the magnetic structure shown in Fig. \ref{Fig1}d.\cite{Curro2000_R6100,Curro2006_173} At higher fields with $B$ applied along the $c$-axis, $B_{int}^{\bot c}(1)$ can be neglected ($B \gg B_{int}^{\bot c}(1)$). The magnetic field along the $c$-axis induces a canting of the Ce local moment\cite{Takeuchi2001_877} (Figure \ref{Fig1}d) leading to extra internal fields $B_{int}^{\| c}(1)$ and $B_{int}^{\| c}(2)$ at both In(1) and In(2) sites, respectively. Therefore the local field at In(1) can be modeled as:
\begin{equation}
B_{local}(1)\simeq \left[1 + K_{c.e.}^{\| c}(1)\right]B + B_{int}^{\| c}(1)
\label{loc01}
\end{equation}
For the Kondo-lattice CeRhIn$_5$, the first term in Eq. (1) is associated with a contribution from itinerant quasiparticles and the second term with the internal field at In(1) due to the out-of-plane Ce-moment component. This internal field component $B_{int}^{\| c}(1) = A_{ord}^{\| c} \mu_{Ce}\cos(\beta/2)$, where $A_{ord}^{\| c}$ is the diagonal $c$-component of the hyperfine coupling tensor from the ordered local moments and $\beta$ is the angle of the conical spin structure (see Fig. \ref{Fig1}d). This term is proportional to $B$ due to the Zeeman interaction. Therefore, the local internal field $B_{local}(1)$ at In(1) sites is:
\begin{equation}
B_{local}(1) = \left[1 + K'(1)\right]B,
\label{loc02}
\end{equation} with $K'(1)$ defining the formal In(1) Knight shift that bears contributions from both local and itinerant spin susceptibilities.
In the case of In(2), the hyperfine field resulting from the in-plane ordered Ce moments follows the oscillatory non-commensurate character of the magnetic structure, $B_{int,0}^{\| c}(2) = B_{i0}$ cos$(2\pi Q_zz)$ (Fig \ref{Fig1}d). The out-of-plane contribution of the Ce moments for the hyperfine field at the In(2) site lies in the $c$-direction\cite{Kambe2007_144433,Ohama2005_094408,Tokunaga2011_214403} and is also proportional to the external field due to the Zeeman interaction. Therefore the local field at an In(2) site can be defined in terms of a formal Knight shift, $K'(2)$:
\begin{equation}
B_{local}(2) = \left[1 + K'(2)\right]B + B_{int,0}^{\| c}(2)
\label{loc03}
\end{equation}

As indicated by solid vertical (gray) arrows in Figs. \ref{Fig1}a and \ref{Fig1}b, below $B^{\ast}$ $\sim$ 30.8 T the position of In(1) transitions can be calculated (see Supplemental Material) assuming a formal Knight shift $K_{\rm{S}}'(1) =$ 7.4(1)\% and quadrupolar frequency $\nu_Q =$ 6.77(1) MHz.

The subscript $\rm{S}$ stands for the magnetic phase AF$_{\rm{S}}$, $B \leq B^{\ast}$, with a small FS and, as introduced later, $\rm{L}$ for AF$_{\rm{L}}$, $B > B^{\ast}$, with a large FS. The formal Knight shift bears contributions from both local and itinerant spin susceptibilities. The value of $K_{\rm{S}}'(1)$ is consistent with the paramagnetic value\cite{Shirer2012_E3067} of $K_{\rm{PM}}^{\| c}(1)\simeq 8.0\%$. The spectrum from In(2) in the AFM phase can be calculated similarly by assuming a periodically oscillating internal field $B_{int,0}^{\| c}(2) = 0.27$ T along the $c$-axis,\cite{Curro2006_173} with nearly the same low-field quadrupolar parameters\cite{Curro2000_R6100,Kohori2000_601} and a formal Knight shift $K_{\rm{S}}'(2) = 1.5(1)$\%. Taking these parameters into account, we calculate the $^{115}$In NMR spectrum that is given by red and green colours for contributions from In(1) and In(2), respectively. The gray solid curve is the simulated (convoluted) overall $^{115}$In NMR spectrum from both In signals.

The simulated spectra in Fig. \ref{Fig1}b are made on the basis of low-field NMR parameters\cite{Curro2000_R6100,Kohori2000_601} that account well for spectra in Fig. \ref{Fig1}a and agree with experiment for fields up to 30.8 T where some deviation from simulation and experimental results begins just where the new AF$_{\rm{L}}$ phase sets in. However, above $B^{\ast} \simeq 30.8$ T, the spectra are well simulated by keeping all low-field nuclear quadrupolar parameters but with an abrupt decrease of both In(1) and In(2) formal shifts from $K_{\rm{S}}'$(1) = 7.4(1)\% to $K_{\rm{L}}'$(1) = 5.1(1)\% and $K_{\rm{S}}'(2) = 1.5(1)$\% to $K_{\rm{L}}'(2) = 1.1(2)$\%, respectively, indicating absence of a detectable local structural distortion at $B^{\ast}$. The simulation remains comparably good at fields well above $B^{\ast}$ (Fig. \ref{Fig1}c). The larger $\Delta K'(1)$ compared to $\Delta K'(2)$ is consistent with the larger hyperfine coupling constant of In(1),\cite{Lin2015_155147} but the relative decrease of $K'$(1) and $K'$(2) is similar, implying a decrease in bulk magnetization\cite{Curro2016_064501} in the high-field state that is reflected in part by a decrease in the slope of the $c$-axis magnetization around $B^{\ast}$.\cite{Takeuchi2001_877} Opening a density-wave gap in the reconstructed large FS is consistent with the decrease in formal shifts if $K_{c.e.}$, which is proportional to the susceptibility of itinerant quasiparticles, dominates $K_{\rm{L}}'$. This is a scenario proposed previously,\cite{Jiao2015_673,Moll2015_6663} but, as we have concluded, the nesting wave vector that opens a gap must be similar to the zero-field $\bm{Q}$. A related scenario is that the decrease in formal Knight shifts is due to a decrease in internal field $B_{int}^{\| c}(1) = A_{ord}^{\| c} \mu_{Ce}\cos(\beta/2)$ that arises from a reduction of the ordered moment, $\mu_{Ce}$, and/or a decrease of the hyperfine coupling constant, $A_{ord}^{\| c}$. Both of these depend on the extent to which Ce's 4$f$ electrons hybridise with band electrons\cite{Curro2016_064501} and, in the limit of stronger hybridisation, would reflect additional $f$ spectral weight being transferred to band states,\cite{Chen2018_066403} with a corresponding increase of the FS. Because a magnetic field tends to weaken Kondo hybridisation as it polarizes spins of both conduction and localized electrons, this scenario superficially seems unlikely but as discussed below is supported by simplified model calculations.

From the high-field data and spectra simulation, we can conclude that the magnetic structure does not change qualitatively through $B^{\ast}$. One possibility is that the magnetic structure adopts the commensurate order with $\bm{Q} = (0.5, 0.5, 0.25)$ observed for CeRhIn$_5$ when $B^{\bot {c}} \gtrsim 2$ T\cite{Raymond2007_242204, Fobes2018_NP} that is not so different from the low-field incommensurate $\bm{Q} = (0.5, 0.5, 0.297)$. For a commensurate $\bm{Q}$, the internal field $B_{int,0}^{\| c}(2)$ at In(2) will take only distinct values, but an incommensurate $\bm{Q}$ creates an oscillating $B_{int,0}^{\| c}(2)$ that produces a characteristic "double horn'' spectral distribution pattern. Such a distribution is, indeed, revealed by the NMR data and simulation presented in Figs. \ref{Fig1}b and \ref{Fig1}c. We conclude that the magnetic structure of CeRhIn$_5$ remains incommensurate with a similar, if not identical, $\bm{Q} = (0.5, 0.5, 0.297)$ above $B^{\ast}$.  

\begin{figure}
\centering
\includegraphics[keepaspectratio,width=8.6cm]{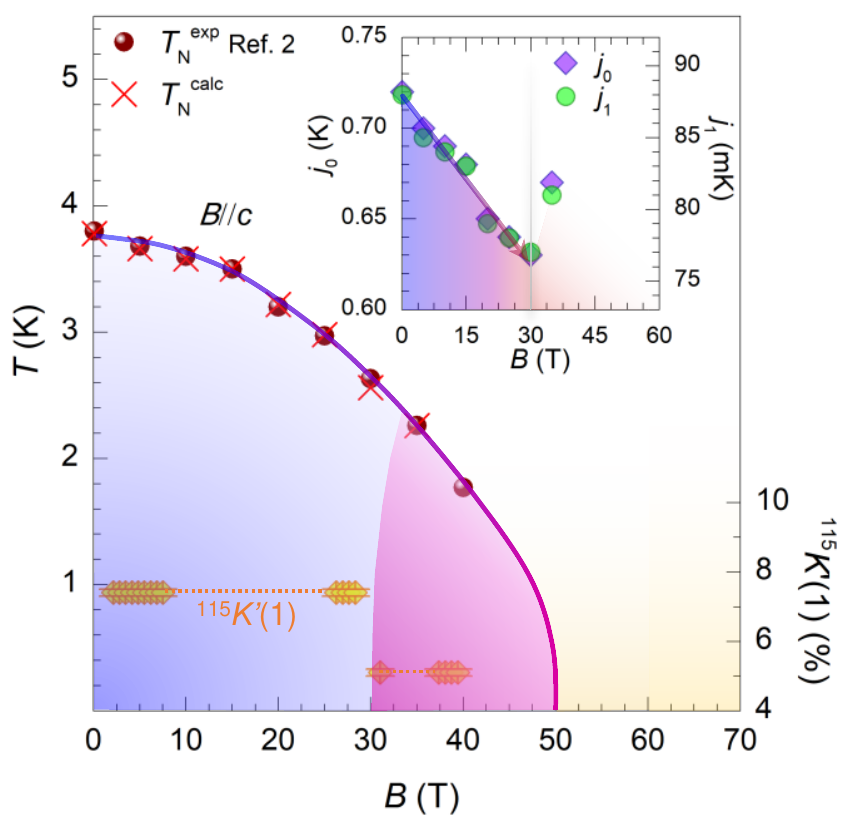}
\caption{Results of a mean-field model with CEF, Zeeman and effective RKKY interactions. Measured and calculated $T_{\rm{N}}(B)$ for the set of coupling constants $j_0$ and $j_1$ shown in the inset. Both coupling constants are suppressed at a similar rate upon increasing magnetic field until $B^{\ast} = 30$ T. Above this field, their values increase, suggesting renormalisation of the effective RKKY interaction due to enhanced Ce $4f$-$c.e.$ hybridisation. See text for a discussion. The yellow diamonds are the measured In(1) formal Knight shift at 0.5 K.}
\label{Fig3}
\end{figure}

At high fields, the In spectrum, acquired in a hybrid 45 T magnet, broadens as shown in Figs. \ref{Fig1}b and \ref{Fig1}c. This broadening is more evident for the equidistant In(1) transitions where the linewidth increases from $\Delta L \simeq$ 0.020(5) T in the low-field limit to $\Delta L \simeq$ 0.10(1) T in the high-field limit. We consider possibilities for this broadening. Though not dramatically, the linewidth increases with increasing fields from 26 to 42 T, which likely is due to the crystal experiencing a slight field gradient in the hybrid magnet. From the magnet's known (in)homogeneity, we estimate that the linewidth would increase by at most 9 \% in this field range. Field-induced electronic anisotropy from the proposed $XY$ nematicity\cite{Ronning2017_313} in principle should contribute to NMR line-broadening. Such a nematic electronic texture would induce anisotropy in the in-plane hyperfine field component at the In(1) site (Fig. \ref{Fig1}d), resulting in line-broadening or even splitting each In(1) transition, and by breaking local tetragonal symmetry of the In(1) site, would produce non-equidistant In(1) transitions due to a modified electric field gradient (EFG). Within the accuracy of our measurements, however, the separation between In(1) transitions remains constant for fields spanning $B^{\ast}$, and there is no clear evidence for splitting of In(1) transitions. Though the pronounced in-plane symmetry breaking of magnetotransport appears at $B^{\ast}$, weak magnetoresistive anisotropy begins to develop\cite{Ronning2017_313} already near 17 T where specific heat and de Haas-van Alphen measurements with field along the $c$-axis also find the onset of enhanced Sommerfeld coefficient and quasiparticle mass.\cite{Jiao2019_045127}  Whether these effects are precursors to proposed nematicity above $B^{\ast}$ is unknown but, whatever their origin, conceivably could manifest in larger linewidths shown in Figs.\ref{Fig1}b and \ref{Fig1}c. Nevertheless, In(1) lineshapes remain symmetric and do not broaden noticeably as field is swept through $B^{\ast}$. The absence of a change in crystal and magnetic structures as a function of field and particularly the abrupt decrease in formal Knight shift at $B^{\ast}$ (Fig. \ref{Fig3}) are primary conclusions that come directly from our NMR measurements.

The ground states of CeRhIn$_5$ and its isostructural family members, CeCoIn$_5$ and CeIrIn$_5$, depend on the orbital character of their 4$f$ wavefunctions that determines the extent of $f$ hybridization with In electronic states.\cite{Willers2015_2384} In a tetragonal environment, the CEF splits the $J$ = 5/2 manifold of CeRhIn$_5$'s 4$f^1$ state into three doublets whose energy separation and wave-functions (see Supplemental Material) have been determined by linear-polarised soft-X-ray absorption and inelastic neutron scattering experiments in zero magnetic field.\cite{Christianson2002_193102,Willers2010_195114} Fields of order $B^{\ast} \simeq 30$ T ($\Delta_{\rm{CEF}}\simeq 7$ meV $\simeq$ 81 K) are sufficient to induce mixing of the wave-functions of the $\Gamma_{7}^{2}$ doublet ground state with the first excited doublet state $\Gamma_{7}^{1}$. This level mixing manifests as a bending of the field-dependent CEF energy levels (see Supplemental Material).

We now consider the consequences of magnetic degrees of freedom. Although a general solution of a theory of a strongly interacting Kondo lattice like CeRhIn$_5$ has not been solved, we incorporate the magnetic Rudderman-Kittel-Kasuya-Yosida (RKKY) interaction into the above electronic framework. This magnetic interaction is represented by an effective spin-spin interaction term, $j_k\mathbf{J_m} \cdot \mathbf{J_n}$. Specifically, we consider a simplified mean-field model with intra- and inter-layer nearest-neighbor (nn) exchange couplings ($j_x = j_y \equiv j_0$ and $j_z \equiv j_1$, Fig. \ref{Fig1}d) to play the role of an effective RKKY interaction combined with the appropriate CEF hamiltonian term (see Supplemental Material).  

Our model does not explicitly include the Kondo interaction but considers it to renormalise the bare spin-spin exchange, so that $j_0$ and $j_1$ are effective exchange coupling constants. With this simple mean-field model we calculate the specific heat thermal dependence (see Supplemental Material) constraining the value of calculated constants to give the zero-field Neel temperature $T_{\rm{N}}$ = 3.8 K and keeping the experimentally determined ratio $j_0/j_1 \simeq 8$.\cite{Das2014_246403} For $B$ = 0, we find effective $j_0$ = 0.72 K and $j_1$ = 0.088 K, which are an order of magnitude smaller than those derived from a model that gives the zero-field magnetic structure.\cite{Das2014_246403} This is consistent to the fact that thermal fluctuations tend to suppress the mean field ordering temperature for a quasi 2D system like CeRhIn$_5$ ($j_0/j_1 \simeq 8$).  

Following the same approach, we estimate the field dependence of $j_0$ and $j_1$, shown in the inset to Fig. \ref{Fig3}, that is required to reproduce the $T_\text{N}(B)$ phase boundary. As seen, $j_0$ and $j_1$ decrease linearly up to 30 T before increasing above $B^{\ast}$. From the Shrieffer-Wolff transformation, the Kondo exchange is proportional to the square of the $f-c.e.$ hybridisation matrix element.\cite{Schrieffer1966_491} A reasonable interpretation of the increase in exchange constants above $B^{\ast}$, then, is that this reflects an enhanced hybrisiation in the high-field state due to field-induced change in the orbital character of the $4f$ wave function. Obviously, a more realistic theoretical framework that explicitly takes into account the Kondo interaction as well as a frustrating inter-layer next nn exchange and orbital degrees of freedom is desirable to substantiate our interpretation.


Our NMR measurements and model calculations thus provide a microscopic basis for the origin of the unusual electronic state that emerges at high fields in the Kondo-lattice CeRhIn$_5$: field-driven mixing of the orbital character of the 4$f$ wave function enhances Kondo hybridisation that induces a large FS above $B^{\ast}$ = 30 T where it experiences a density-wave instability due to nesting at a $\bm{Q}$ close to, if not the same as, that characterising magnetic order in the zero-field antiferromagnetic state. There is no detectable change in local structure at fields to 42 T. Except for the field scale $B^{\ast}$, which is specific to the Kondo interaction and crystal-field wave functions of CeRhIn$_5$, similar high-field states should be generic to Kondo-lattice materials. With the essential role of the orbital nature of wave functions and its consequences for Kondo coupling, $B^{\ast}$ could be considered in the zero-temperature limit to reflect an orbitally selective type of Kondo-breakdown quantum-critical point\cite{Pepin2007_206401,Si2001_804} within the ordered state. This is an interpretation suggested initially by Jiao \textit{et al.}\cite{Jiao2015_673} and now we provide a microscopic rationale for it.

\begin{acknowledgments}
We acknowledge enlightening discussions with F. Ronning, P. F. S. Rosa, M. Smidman, L. Jiao, H. Q. Yuan, D. J. Garcia, N. Curro and C. Rettori. Work at State University of Campinas (Unicamp) was supported by CNPq ($\#$ 307668/2015-0) and FAPESP through grants $\#$ 2012/05903-6 and 2012/04870-7 and at Los Alamos by the U.S. Department of Energy, Division of Materials Sciences and Engineering. Part of this work was performed and supported by JSPS KAKENHI through grant number JP16KK0106 and by REIMEI Research Program of JAEA. The work at the National High Magnetic Field Laboratory was supported by the National Science Foundation Cooperative Agreement No. DMR-1157490 and the State of Florida.
\end{acknowledgments}

\FloatBarrier



\end{document}


\title{Supplemental material for:\\Orbitally defined field-induced electronic state in a Kondo lattice}

\author{G. G. Lesseux}
\affiliation{Instituto de F\'{\i}sica "Gleb Wataghin", Universidade Estadual de Campinas, 13083-859, Campinas, SP, Brazil}

\author{H. Sakai}
\affiliation{Advanced Science Research Center, Japan Atomic Energy Agency, Tokai, Ibaraki, 319-1195, Japan}

\author{T. Hattori}
\affiliation{Advanced Science Research Center, Japan Atomic Energy Agency, Tokai, Ibaraki, 319-1195, Japan}

\author{Y. Tokunaga}
\affiliation{Advanced Science Research Center, Japan Atomic Energy Agency, Tokai, Ibaraki, 319-1195, Japan}

\author{S. Kambe}
\affiliation{Advanced Science Research Center, Japan Atomic Energy Agency, Tokai, Ibaraki, 319-1195, Japan}

\author{P. L. Kuhns}
\affiliation{National High Magnetic Field Laboratory, Florida State University, Tallahassee-FL, 32310, U.S.A.}

\author{ A. P. Reyes}
\affiliation{National High Magnetic Field Laboratory, Florida State University, Tallahassee-FL, 32310, U.S.A.}

\author{J. D. Thompson}
\affiliation{Los Alamos National Laboratory, 87545, Los Alamos, NM, USA}

\author{P. G. Pagliuso}
\affiliation{Instituto de F\'{\i}sica "Gleb Wataghin", Universidade Estadual de Campinas, 13083-859, Campinas, SP, Brazil}

\author{R. R. Urbano}
\affiliation{Instituto de F\'{\i}sica "Gleb Wataghin", Universidade Estadual de Campinas, 13083-859, Campinas, SP, Brazil}

\maketitle

\section*{1. $^{115}$In NMR in CeRhIn$_5$}

The field-swept $^{115}$In-NMR spectra (In nuclear spin $I$ = $9/2$, natural abundance 95.7$\%$ and nuclear gyromagnetic ratio $^{115}\gamma/2\pi$ $=$ 9.3295 MHz$\cdot$T$^{-1}$) were obtained at 500 mK and at constant frequencies of 48.5 MHz (~ 5.2 T), 290.65 MHz (~ 31.2 T) and 393.6 MHz (~ 42.2 T).
A nucleus with $I$ $\geq$ 1 such as $^{115}$In has both an electric quadrupole moment $eQ$ and a magnetic dipole moment. In this situation, for non-cubic sites the degeneracy of nuclear energy levels is lifted even at zero magnetic field due to the interaction between $eQ$ and the electric field gradient (EFG) described by:

\begin{equation}
{\cal H}_Q = \frac{h\nu_Q}{2}\left[3I^{2}_{z}-I^2+\frac{\eta}{2}\left(I_{+}^{2}+I_{-}^{^2}\right)\right]
\label{EFG}
\end{equation}

where $h$ is Planck's constant, $\nu_Q$ $=$ $e^2qQ/2I(2I-1)$ is the quadrupole frequency along the principal axis of the EFG, with $eq = V_{zz}$. The asymmetry parameter of the EFG is defined as $eta$ $=$ $(V_{xx}-V_{yy})/V_{zz}$ with the second derivative of the electrical potential $V_{\alpha\alpha}$ being the EFG along the direction $\alpha$ ($\alpha = x,y,z$).

When $^{115}$In experiences an EFG, the ten degenerate nuclear spin states split into five energy levels, yielding four resonance frequencies. However, for the NMR measurements, we applied magnetic fields to lift the degeneracy of the magnetic dipole degrees of freedom, even though the nuclear energy levels were already split by the electric quadrupole interaction. Therefore, the total effective Hamiltonian is described by:

\begin{equation}
\begin{split}
{\cal H}_N &= {\cal H}_Q + {\cal H}_Z = \frac{h\nu_Q}{2}\left[3I^{2}_{z}-I^2+\frac{\eta}{2}\left(I_{+}^{2}+I_{-}^{^2}\right)\right]
+\hbar\gamma_n\left[1 + K'(2)\right]I\cdot B
\end{split}
\label{Htotal}
\end{equation}
where $K$ is commonly defined as Knight shift, $B$ is the applied magnetic field and the term $(1+K)B$ is defined as the effective field $B_{eff}$.

In the following section, the Knight shift is redefined for the particular case, here as a formal Knight shift $K'$, by taking into account the internal field created by the Ce 4$f$ magnetic moments at the In sites.

\newpage

\section*{2. Simulation of high-field $^{115}$In NMR spectra}

The In(2) spectra in the ordered phase ($T$ = 0.5 K) were simulated by assuming an oscillating internal field $z$-component given by $B_{int}^{\| c}(2) = B_{i0}\cos{(2\pi Qz)}$, where $B_{i0}$ = 0.27 T was estimated at zero field \cite{Curro2000_R6100,Kohori2000_601} and $Q_z$ is the $z$-component of the incommensurate AF wave-vector $\bf{Q}$.

The fields of resonance for a given frequency were calculated after diagonalising the effective nuclear spin Hamiltonian of Eq. \ref{Htotal} assuming:
\begin{equation}
B_{eff} = [1+K'(2)]B + B_{int}^{\| c}(2) 
\label{Beff}
\end{equation}

Hence, the NMR pattern was obtained from a convolution of the calculated fields of the resonance with a Lorentzian function. It is worth mentioning that any incommensurate value of $Q_z$ will lead to the same double horn pattern when the magnetic field is along the c-axis. Although the $Q_z$-value is undetermined based solely on these NMR results, the In(2) NMR spectrum is very sensitive to a transition from incommensurate to commensurate $Q_z$.

\newpage

\section*{3. Field-induced evolution of Ce's 4\textit{f} orbitals}

\begin{figure}[b]
\includegraphics[keepaspectratio,width=8.6cm]{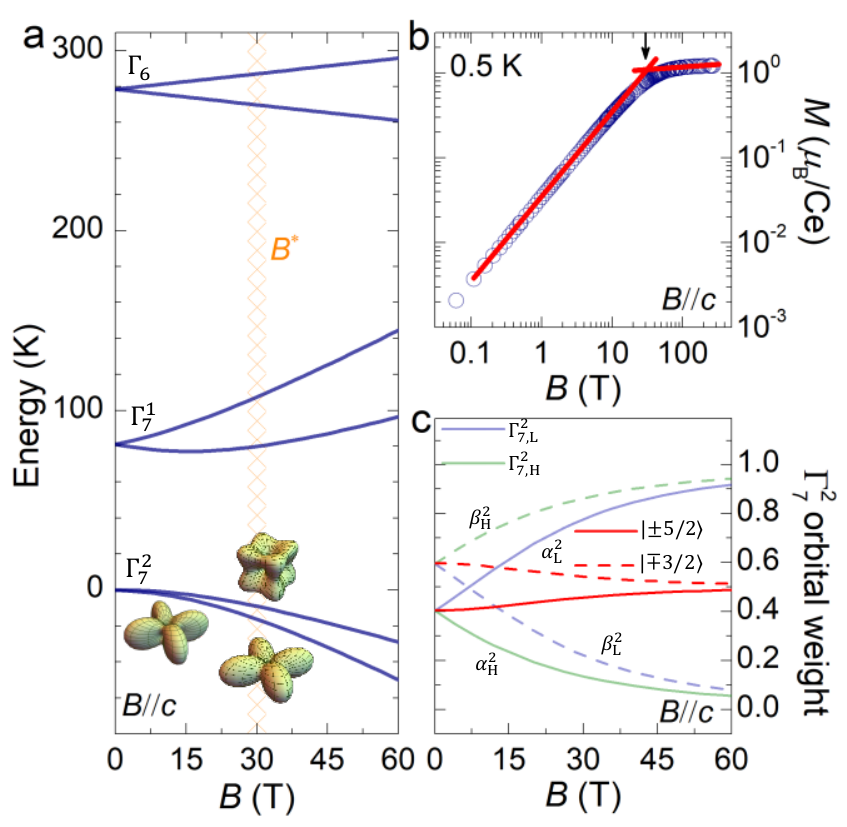}
\caption{${4f}$ crystal-electric field response to magnetic field. \textbf{a} Field-dependent Ce $4f$ CEF energy level scheme for $B\|c$ (solid blue lines). The hashed orange area near 30 T represents the field range where mixing between the ground- and first-excited $\Gamma_7$ CEF doublets occurs for $B\|c$. This mixing is reflected as well by a change in slope of magnetisation (\textbf{b}) and occupancy of the pure $\vert\pm 5/2\rangle$\ (\textbf{c}) near 30 T. \textbf{b} Simulation of the magnetisation as a function of $B\|c$ considering only CEF and Zeeman interactions at 0.5 K. This simulation produces a weaker slope but continued growth of $M(H)$ above 30 T. The extrapolated solid red lines through calculated points indicate a crossover point between orbital states at 30 T. \textbf{c} Field dependence of the zero-field $\Gamma_{7}^{2}$ CEF doublet ground-state: the field-split eigenstates $\Gamma_{7,H}^{2}$ $= \alpha_H\vert+5/2\rangle - \beta_H\vert-3/2\rangle$ (green color) and $\Gamma_{7,L}^{2}$ $= \alpha_L\vert-5/2\rangle - \beta_L\vert+3/2\rangle$ (blue color) and individual field-dependent orbital weight contributions $\vert\pm 3/2\rangle$ (dashed lines) and $\vert\pm 5/2\rangle$ (solid lines) of $\Gamma_{7,H}^{2}$ and $\Gamma_{7,L}^{2}$ as well as the averaged orbital weight contributions of $\vert\pm 5/2\rangle$ (solid red line) and $\vert\pm 3/2\rangle$ (dashed red line). The red solid line (\textit{dashed line}) indicates an overall enhancement (\textit{suppression}) of the more oblate $\vert\pm 5/2\rangle$ (prolate $\vert\pm 3/2\rangle$) orbital upon increasing magnetic field. The evolution of the spatial form of $\Gamma_{7}^{2}$ orbital in real space for zero-field and $B^{\|c} = 30 $T (field split $\Gamma_{7,L}^{2}$ and $\Gamma_{7,H}^{2}$) is illustrated in \textbf{a}.}
\label{Fig2}
\end{figure}

Starting with CEF parameters \cite{Willers2010_195114} of the zero-field doublet wave functions $\Gamma_{7}^{2}$, $\Gamma_{7}^{1}$ and $\Gamma_{6}$ and fully diagonalizing the CEF + Zeeman Hamiltonian \cite{Pagliuso2006_08P703}, we calculate the effect of a magnetic field on these levels. The solid blue lines in Fig. \ref{Fig2}A represent the field-dependent CEF energy levels of CeRhIn$_5$ for $B\|c$. 

Fields of order $B^{\ast} \simeq 30$ T ($\Delta_{\rm{CEF}}\simeq 7$ meV $\simeq$ 81 K) are sufficient to induce mixing of the wave-functions of the $\Gamma_{7}^{2}$ doublet ground state with the first excited doublet state $\Gamma_{7}^{1}$. This level mixing manifests as a bending of the field-dependent CEF energy levels as shown in Fig. \ref{Fig2}A for $B\|c$. Ignoring possible magnetic interactions (and consequently magnetic order) as well a Kondo hybridization that broadens the the ground and excited CEF levels, we calculate the corresponding magnetisation as a function of applied magnetic field $B\|c$ at 0.5 K, with the result shown in Fig. \ref{Fig2}B. The calculated magnetisation bends over to a region of weaker yet monotonically increasing field dependence above 30 T, which is consistent with the decrease in formal Knight shift and enhanced hybridization near $B^{\ast}$ as well as direct measurements of the magnetization.\cite{Takeuchi2001_877} Given the simplicity of this approach, we do not expect the calculated field-dependent magnetisation to capture all features of the experimental data, but the result is suggestive.

In Fig. \ref{Fig2}C we present the field dependence of the weights of pure $\vert\pm 5/2 \rangle$ and $\vert\mp 3/2 \rangle$ orbitals. The zero-field CEF $\Gamma_{7}^{2}$ doublet ground-state of CeRhIn$_5$ is a linear combination of both orbitals $\Gamma_{7}^{2} \equiv \vert 0 \rangle = \alpha \vert \pm 5/2 \rangle - \beta \vert \mp 3/2 \rangle$ with $\alpha = 0.62$ and $\beta = 0.78$.\cite{Christianson2002_193102,Willers2010_195114} As seen in this figure, the orbital contribution (weight) from the pure $\vert\mp 3/2 \rangle$ gradually decreases and that from the pure $\vert\pm 5/2 \rangle$ grows with increasing field. With its in-plane oblate configuration, the stronger $\vert\pm 5/2 \rangle$ character of occupied Ce $f$ orbitals at high fields promotes $f$-$c.e.$ hybridisation between Ce and In(1) orbitals when $B\|c$, even though field-induced polarization of spins tends to weaken hybridization. The Kondo effect and mixing of orbital contributions to the ground state should be smooth functions of applied field, but these results suggest that the competition between apposing tendencies is tipped abruptly in favor of increased hybridisation near $B^{\ast}$ and is the origin of anomalies found in various physical properties. Illustrations of the actual shapes of the overall $\Gamma_{7}^{2}$ ground-states orbitals in real space at zero-field and 30 T are displayed in Figure \ref{Fig2}A.

\newpage

\section*{4. Calculation of the mean field specific heat}
The calculation of the specific heat data was obtained by solving the Hamiltonian used to describe the RKKY and CEF interactions in CeRhIn$_5$ given by:\cite{Pagliuso2006_08P703}

\begin{equation}
\begin{split}
{\cal H} = \sum_{\langle m,n \rangle}j_k\mathbf{J_m} \cdot \mathbf{J_n} \label{H_MF}
+ \sum_m \left[B_{20}\mathbf{O^0_{2,m}} + B_{40}\mathbf{O^0_{4,m}} + B_{44}\mathbf{O^4_{4,m}}\right]
+ g_J\mu_B\mathbf{J\cdot B}. 
\end{split}
\end{equation} where $j_k$ (with $k$ = 0 or 1, as indicated in Fig. 1d in the main text) represents the effective RKKY exchange interactions between neighbor spins, $\mathbf{J_m}$ and $\mathbf{J_n}$, and the second and third terms are the tetragonal CEF and Zeeman contributions, respectively. The on-site crystal field interaction is therefore fully taken into account. This model does not explicitly include the Kondo interaction but considers it to renormalise the bare spin-spin exchange, so that $j_0$ and $j_1$ in Eq. \ref{H_MF} are effective exchange coupling constants. The first term in Eq. \ref{H_MF}, then, "mimics'' magnetic (spin) interactions in a generic Kondo lattice but is unable to predict the magnetic structure of CeRhIn$_5$ that requires the inclusion of next-nearest neighbor interactions.\cite{Fobes2018_NP}

\begin{figure*}[b]
\centering
\includegraphics[width=12.9cm]{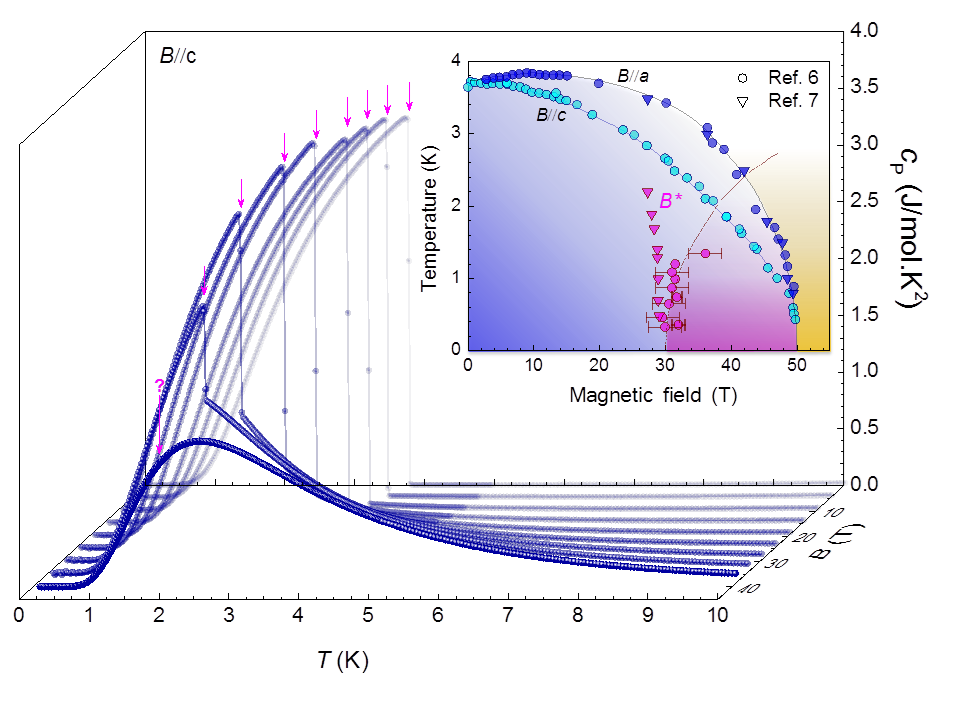}
\caption{Specific heat from a mean-field model with CEF, Zeeman and effective RKKY interactions. Calculated mean-field specific heat for several magnetic fields applied along the $c$-axis at low temperatures. The red arrows indicate the measured magnetic ordering transition temperatures $T_{\rm{N}}(B)$.\cite{Jiao2015_673} This model does not predict long-range magnetic order for $B\geq 40$ T, $j_0 \leq$ 0.72 K and $j_1 \leq$ 0.088 K, but only gives a Schottky anomaly indicated by a question mark. This also signals the influence of interactions/fluctuations not accounted by this framework, specially for $B \geq B^{\ast}$.The inset shows temperature versus magnetic field phase diagram for CeRhIn$_5$. The circles and triangles are data extracted from Refs. \onlinecite{Jiao2015_673} and \onlinecite{Ronning2017_313}, respectively.}
\label{SI1}
\end{figure*}

With a mean field approximation where $\mathbf{J_m}\cdot \mathbf{J_n}$ $\sim$ $\mathbf{J_m}\left\langle \mathbf{J_n}\right\rangle$, we numerically diagonalised the electronic Hamiltonian of eq. \ref{H_MF} for each increment of applied magnetic field and evaluated the eigenvalues $E_i$ and corresponding eigenfunctions:
\begin{equation}
\left|\phi_i \right\rangle=\sum_{m=-\texttt{J}}^{\texttt{J}} c_{im}\left|m\right\rangle
\label{eig_fun_MF}
\end{equation} where $\left|m\right\rangle$ form a complete basis for the manifold of angular momentum $\mathbf{J}$. The internal ($U$) and Helmholtz ($F$) free energies were calculated using the equations:
\begin{align}
U &= \frac{1}{Z}\sum_{i}E_i \texttt{exp}\left(E_i/k_{B}T\right)\\
F &= -k_{B}T\texttt{ln}Z
\end{align} where $Z=\sum_{i}\texttt{exp}\left(E_i/k_{B}T\right)$ is the canonical partition function and $k_B$ is the Boltzmann constant. Therefore, the specific heat $c$ is finally obtained by:
\begin{equation}
c = T\frac{\partial S}{\partial T}
\end{equation} after computing the entropy $S$ by using the relation $F = U - TS$.

Hence, Figure \ref{SI1} shows the magnetic field dependence of the specific heat as a function of temperature simulated to reproduce the measured $B$ vs $T$ phase diagram displayed in Fig. 3 in the main text. The crystal field parameters used for this calculation, $B_{20} =$ -10.789 K, $B_{40} =$ 0.60343 K and $B_{44} =$ 1.4854 K, were extracted from Ref. \onlinecite{Willers2010_195114}. The Kondo interaction as well as the fluctuations were not taken into account by this mean field framework. Nonetheless, it was still used to qualitatively evaluate the attenuation of effective exchange parameters as a function of magnetic field.

\FloatBarrier
